\documentclass[twocolumn,pre,tightenlines,nofootinbib,superscriptaddress,showpacs]{revtex4-1}
\usepackage{amsmath}
\usepackage{amssymb,amsfonts,latexsym}
\usepackage{bm}
\usepackage[mathcal]{euscript}
\usepackage{graphicx}
\usepackage{epsfig}
\usepackage{color}

\newcommand{\beq}{\begin{equation}}
\newcommand{\eeq}{\end{equation}}

\newcommand{\eb}[1]{{\color{black} #1}}

\begin{document}

\title{Comparison between Smoluchowski and Boltzmann approaches for
self-propelled rods}

\author{Eric Bertin}
\affiliation{Universit\'e Grenoble Alpes, LIPHY, F-38000 Grenoble, France}
\affiliation{CNRS, LIPHY, F-38000 Grenoble, France}

\author{Aparna Baskaran}
\affiliation{Department of Physics, Brandeis University, Waltham MA 02474, USA}

\author{Hugues Chat\'e}
\affiliation{Service de Physique de l'Etat Condens\'e, CNRS UMR 3680, CEA-Saclay, 91191 Gif-sur-Yvette, France}
\affiliation{Beijing Computational Science Research Center, Beijing 100094, China}

\author{M. Cristina Marchetti}
\affiliation{Department of Physics, Syracuse University, Syracuse NY 13244, USA}
\affiliation{Syracuse Biomaterials Institute, Syracuse University, Syracuse NY 13244, USA}

\begin{abstract}
Considering systems of self-propelled polar particles with nematic
interactions (``rods''), we compare the continuum equations describing the
evolution of polar and nematic order parameters, derived either from
Smoluchowski or Boltzmann equations. Our main goal is to understand the
discrepancies between the continuum equations obtained so far in both
frameworks. We first show that in the simple case of point-like particles
with only alignment interactions, the continuum equations obtained have the
same structure in both cases. We further study, in the
Smoluchowski framework, the case where an interaction force is added on top
of the aligning torque. This clarifies the origin of the additional terms
obtained in previous works. Our observations lead us to emphasize the need for a more involved
closure scheme than the standard normal form of the distribution when dealing with active systems.
\end{abstract}

\maketitle

\section{Introduction}

``Self-propelled rods", {\it i.e.} elongated objects spending energy to displace themselves typically along their long axis,
are among the most generic and ubiquitous objects studied in active matter physics.
Living and inert examples abound: elongated bacteria swimming or crawling on a surface \cite{BACTERIA}, 
chemically-propelled micro- and nano-rods \cite{NANORODS}, biofilaments displaced by molecular motors \cite{Schaller10,Schaller11a,Schaller11b,Sumino,Sanchez}, 
shaken granular particles \cite{Sood}, etc.
Their main interaction often amounts to alignment due, e.g., to direct contact via  collisions. 
In many cases  the fluid in which the self-propelled rods move can be treated as inert, simply providing friction. This ``dry active matter" setting nevertheless
shows non-trivial collective properties \cite{Peruani06,Ginelli10,Wensink12,McCandlish12,Menzel13,Gao15,Kuan}.
Among these, the emergence of dense (almost close packed), polarly-oriented clusters has been noted repeatedly
and is a quite natural outcome of aligning collisions: when two rods meet, they often either end up aligned or anti-aligned (``nematic" alignment).
In the first case, they will stay alongside each other for quite a while, even in the absence of attractive interactions,
and may recruit others, forming clusters.

Numerical work on various models of moving elongated objects interacting via  steric repulsion
has revealed the possibility of complex (and not yet fully understood) phase diagrams, even for simple rigid rods.
 The ubiquity of dense clusters, forming even at low global densities,
seems a priori to be a major difficulty in building
theoretical, continuum descriptions of these systems since most approaches so far rely on the hypotheses
that only binary interactions need to be considered and/or that rods decorrelate between collision events.
Nevertheless, such continuum descriptions have been considered, notably by us, in the past.

Theoretical approaches typically yield continuum (or ``hydrodynamic'') equations governing a nematic and a polar order field, as well as
a continuity equation for the density field. 
\eb{By 'continuum equations', we mean here a reduced set of evolution equations for conserved and order parameter fields.}
The first derivations of such sets of equations were performed by Baskaran and Marchetti, who treated
explicit collisions between thin rods \cite{BaskaranPRE08,BaskaranPRL08}, while a different route, treating aligning \eb{point-like} particles, was followed by Peshkov {\it et al.} \cite{Peshkov12}.
These works predicted the emergence of global nematic order at the deterministic level, that is, without including noise in the continuum equations.

Baskaran and Marchetti first studied the case of interacting self-propelled rods assuming overdamped microscopic dynamics from the outset \cite{BaskaranPRE08}. In a second paper they included inertia in the  microscopic dynamics \cite{BaskaranPRL08}, with linear and angular momentum transfer during collisions between rods. They derived a Fokker-Planck equation for the joint probability distribution of positions and velocity, and then took the overdamped limit of the kinetic equation, obtaining a Smoluchowski equation for the one-particle probability density. Continuum equations for the density, polar and nematic fields were then obtained from the Smoluchowski equation, and the linear stability of the basic homogeneous states was studied.  This derivation yields important modifications of the Smoluchowski equation that result in nonlinearities in the continuum equations that are not obtained when considering overdamped dynamics \eb{from} the outset.  Here we focus on the equations reported in \cite{BaskaranPRL08}.

Following the kinetic approach pioneered by \cite{BDG} for polar constant-speed \eb{point-like} particles aligning ferromagnetically,
Peshkov {\it et al.}~treated the case where the same particles align nematically, i.e.~anti-align when their incoming angle is larger than $\frac{\pi}{2}$ \cite{Peshkov12}.
They obtained well-behaved nonlinear partial differential equations whose solutions were shown to be in good qualitative agreement with the Vicsek-like
model they were derived from.

The sets of equations resulting from these works bear strong similarities, but also  differences.
In this paper, we explain and discuss the origin of these differences.

\section{Models and main goals}
\label{sect-maingoals}

We consider self-propelled polar particles moving in two dimensions at constant speed 
$v_0$. When isolated, their polarity angle $\theta$ diffuses at long times. They align
nematically with neighbors. To match
previous studies \cite{BaskaranPRE08,BaskaranPRL08,Peshkov12,Peshkov14}, we
consider either pure diffusion of the angle $\theta$ with rotational
diffusion coefficient $D_R$, or 'run-and-tumble' type dynamics where $%
\theta$ is changed, with a probability $\lambda_R$ per unit time, into $%
\theta^{\prime }=\theta+\eta$, where $\eta$ is a random variable of
distribution $P(\eta)$. Moreover, we neglect positional diffusion
in sections~\ref{sect-maingoals} and \ref{sect-point-like}, and
later reintroduce it in section~\ref{sect-force}.

In the following, we compare continuum equations obtained starting from a
Smoluchowski equation and from a Boltzmann equation. Both these kinetic equations
govern the time evolution of the probability $f(\mathbf{r}%
,\theta ,t)$ to have a particle at position $\mathbf{r}$ with polarity $%
\theta $, at time $t$. 
Although derived from different starting points, these
equations bear strong formal similarities, and one may thus
expect the resulting continuum equations describing the evolution of the
relevant order parameter fields (here, the polar and nematic fields) to be
similar, with changes affecting only the precise values of the coefficients.
It was found, however, that the continuum equations obtained in
\cite{BaskaranPRE08,BaskaranPRL08} starting from the Smoluchowski equation
on one side, and in \cite{Peshkov12,Peshkov14} starting from the Boltzmann
equation on the other side, possess many differences, as described
below. The goal of this paper is on the one hand to emphasize the formal
similarities of the two approaches, by formulating them in a common
framework, and on the other hand to outline the origins of the differences in the
resulting continuum equations.

\subsection{Comparing Smoluchowski and Boltzmann equations}

The Smoluchowski and Boltzmann equations for the probability density $f(\mathbf{r},\theta ,t)$ are respectively given by
% (neglecting positional diffusion),  %
\begin{equation}
\partial _{t}f+v_0\mathbf{e}(\theta )\cdot\bm \nabla f=D_{R}\partial _{\theta
}^{2}f-\partial _{\theta }(f\tau )-{\bm\nabla }\cdot (f{\bm F})  \label{1.1}
\end{equation}%
\begin{equation}
\partial _{t}f+v_0\mathbf{e}(\theta )\cdot\bm \nabla f= I_{\mathrm{dif}}[f]
+I_{\mathrm{col}}[f]  \label{1.2}
\end{equation}%
(see below for notations).
The Smoluchowski equation~\eqref{1.1} was derived in \cite{BaskaranJSM10} by coarse-graining a 
microscopic model of long, hard thin rods undergoing collisions. 
That work also included translational diffusion, which is neglected here to simplify the comparison between the two approaches 
(the effect of translational diffusion is discussed in Sec.~\ref{sect-force}).
The Boltzmann equation~\eqref{1.2} was obtained in \cite{Peshkov12} for point particles with prescribed nematic alignment interactions.
On the left hand side both equations contain a convective mass flux due to self-propulsion at speed $v_0$, with  $\mathbf{e}(\theta )=\left(\cos\theta,\sin\theta\right)$
a unit vector along the direction of self-propulsion. The speed $v_0$ was set equal to $1$ in \cite{Peshkov12} (without loss of generality), but is retained here for clarity.

The content of the right hand side of Eq. (\ref{1.1}) can be described as
follows: $D_{R}$ is a rotational diffusion coefficient representing a
Gaussian white noise process that reorients the particle's self-propulsion
velocity at each time step; $\tau $  is the mean-field torque
exerted by the other rods, given by (to first order in spatial gradients)
\begin{eqnarray}
\tau(\mathbf{r},\theta,t) &=&\int d\theta ^{\prime }K_{1}\left( \theta -\theta
^{\prime }\right) f\left( \mathbf{r},\theta ^{\prime },t\right) \notag\\
&&
+ \int d\theta ^{\prime } \, \mathbf{K}_{2}(\theta^{\prime },\theta) \cdot \bm\nabla f\left( \mathbf{r},\theta ^{\prime },t\right) \;,
\label{eq:tau}
\end{eqnarray}
where $K_{1}$ and $\mathbf{K}_{2}$ are collision kernels describing
hard rods collisions. The second term in Eq.~\eqref{eq:tau} takes into account the difference in position of the center of the colliding rods due to their finite size (see Appendix~\ref{appendix-A}). Finally,  the mean-field force $\mathbf{F}$ is given by (to lowest order in gradients)
\begin{equation}
\label{eq:F}
\mathbf{F}=\int_{-\pi }^{\pi }d\theta ^{\prime }\mathbf{G}(\theta ,\theta ^{\prime })f(%
\mathbf{r},\theta ^{\prime },t)\;,
\end{equation}
where $\mathbf{G}$ is again a hard rod
collision kernel describing linear momentum
transfer in a collision.

The content of the right hand side of Eq.~(\ref{1.2}) can be described as
follows:  $I_{\mathrm{dif}}[f]$ is a generator of rotational reorientation
through an arbitrary stochastic process $\eta $,
\begin{equation}
I_{\mathrm{dif}}[f] = \lambda_R \int_{-\pi }^{\pi }d\theta ^{\prime }f(\theta
^{\prime })\big[\langle \delta_{2\pi }(\theta ^{\prime }+\eta -\theta
)\rangle _{\eta }-\delta_{2\pi } (\theta ^{\prime }-\theta )\big]  \label{1.3}
\end{equation}%
(with $\lambda_R$ a frequency of ``tumbling'' events)
and $I_{\mathrm{col}}[f]$ is the collision integral defined as
\begin{eqnarray}
I_{\mathrm{col}}[f] &=&\int_{-\pi }^{\pi }d\theta _{1}\int_{-\pi }^{\pi
}d\theta _{2}K_{B}(\theta _{2}-\theta _{1})f(\mathbf{r},\theta _{1})f(%
\mathbf{r},\theta _{2})  \notag \\
&\times &\Big[\langle \delta _{2\pi }\big(\Psi (\theta _{1},\theta
_{2})+\eta -\theta \big)\rangle _{\eta }-\delta _{2\pi }(\theta _{1}-\theta )%
\Big]  \label{1.4}
\end{eqnarray}%
where $\delta_{2\pi }$ is a generalized Dirac distribution taking into account the $2\pi$-periodicity of angles.
The notation $\langle \dots \rangle _{\eta }$ indicates an average over the
noise distribution $P(\eta )$, and $\Psi (\theta _{1},\theta _{2})$ is the
direction of motion of particle $1$ after the collision between particles $1$ and $2$, up to an additive noise $\eta$ (see Sect.~\ref{sect-point-like} for details).
We assume here that $\Psi (\theta _{1},\theta _{2})$ favors nematic
alignment.

There are three differences between these two kinetic equations.
\eb{(i)} First, the reorientation events in the
Boltzmann approach are described by an
arbitrary stochastic process with probability distribution $P\left( \eta \right)$.
For a uniform distribution $P(\eta)$ over $[-\pi,\pi]$, one recovers run-and-tumble dynamics,
while a (wrapped) Gaussian $P\left( \eta \right)$ yields Brownian rotational diffusion at rate $\lambda_R$.
In the Smoluchowski equation, in contrast, one has assumed Gaussian rotational noise from the outset. This slight difference, 
however, has no influence on the structure of the resulting continuum equations.
Moreover, when a van Kampen expansion
is carried out in Eq. (\ref{1.3}), in the case of a distribution $P(\eta)$ with a small variance $\sigma^2 \ll 1$, one finds
$I_{\mathrm{dif}}[f]\simeq D_R \partial_{\theta }^{2}f$, with $D_R=\lambda_R \sigma^2$,
so that one recovers a rotational diffusion term as in Eq.~\eqref{1.1}.
\eb{(ii)} Secondly, the Smoluchowski equation contains spatial gradients  in both force and torque, while the collision integral is local in the Boltzmann equation.
This difference arises because the Smoluchowski equation describes rods of finite length $l$, necessitating an additional gradient expansion in the size of
the particles,  while the Boltzmann equation assumes point
particles. 
\eb{(iii)} Finally, a third subtle distinction exists between the two theories and lies in
the details of the collision kernels whose explicit forms are discussed below. In the Boltzmann description, combining the prescribed nematic alignment rule with the kinetics of collision leads to a kernel of mixed polar and nematic symmetry. On the other hand the collision kernel $K_1$ associated to the torque in
the Smoluchowski equation, calculated from hard rod collisions \cite{BaskaranPRL08}, has nematic symmetry in the limit of infinitely thin rods. As we shall show below, the nematic symmetry is also recovered in the Boltzmann case by considering the limit of infinitely thin rods (while the opposite limit of quasi-circular particles was originally considered in \cite{Peshkov12}).

\subsection{Comparing the derived macroscopic hydrodynamics}

Macroscopic continuum equations have been
derived from both the above Smoluchowski and Boltzmann equations in \cite{BaskaranPRL08} and in  \cite{Peshkov12,Peshkov14}.
A first step consists in transforming these equations into a hierarchy of field equations by introducing the Fourier expansion of  $f(\mathbf{r},\theta ,t)$ in $\theta$:
\begin{subequations}
\begin{gather}
f_k(\bm r, t) = \int_{-\pi}^{\pi} d\theta f(\bm r, \theta,t)\, e^{ik\theta} \\
f(\bm r, \theta,t) = \frac{1}{2\pi}\sum_{k=-\infty}^{\infty} f_k(\bm r, t)\, e^{-ik\theta}
\end{gather}
\end{subequations}

The complex fields $f_{1}$ and $f_{2}$ are related to the vectorial polar field ${\bf P}$ and to the tensorial nematic field ${\bf Q}$ as
\begin{equation}
\rho {\bf P} \!=\! \left(\!\begin{array}{lr}
{\rm Re}f_1\\
{\rm Im}f_1 \end{array} \!\right) , \;
\rho {\bf Q} \!=\! \frac{1}{2} \!\left(\begin{array}{lr}
{\rm Re}f_2 & {\rm Im}f_2 \\
{\rm Im}f_2 & -{\rm Re}f_2\end{array} \!\right) \;,
\end{equation}
where $\rho=f_0$ denotes the density field. 
To compare the continuum equations obtained in \cite{BaskaranPRL08} and in  \cite{Peshkov12,Peshkov14}, 
we rewrite them in a common notation. In two dimensions, the
complex notation $f_{1}$ and $f_{2}$ is much more convenient than the more standard vectorial/tensorial notations.

The continuum equations obtained in
\cite{BaskaranPRL08} read (neglecting terms arising from spatial diffusion for
ease of comparison)\footnote{In \cite{BaskaranPRL08}, a 'renormalized' speed $\tilde{v_0}$ has been found in the equation for $f_1$ instead of $v_0$, but we neglect this correction at this stage to simplify the discussion. We come back to this point in Sec.~\ref{sect-force}.}
\begin{subequations}
\label{eq:SmoluBM08}
\begin{align}
\partial _{t}\rho &+v_{0}\mathrm{Re}(\nabla ^{\ast }f_{1})=0\;,
\label{eq:rho-SmoluBM08}\\
\partial _{t}f_{1} &= -D_{R}f_{1}+\zeta_S f_{1}^{\ast }f_{2}-\frac{v_{0}}{2}\left(\nabla \rho +\nabla ^{\ast }f_{2}\right) \notag \\
&-\lambda ^{\prime }\big(f_{1}^{\ast }\nabla f_{1}+f_{1}\nabla ^{\ast}f_{1}
-f_{1}\nabla f_{1}^{\ast }\big)\;,  
\label{eq:f1-SmoluBM08} \\
\partial _{t}f_{2}&=\mu _{S}f_{2}-\frac{v_{0}}{2}\nabla f_{1}
-\frac{\kappa_S }{2}f_{1}^{\ast }\nabla
f_{2}-\frac{\kappa_S^{\prime} }{2}f_{2}\nabla f_{1}^{\ast}
\notag \\
&-\frac{\chi_S }{2}%
f_2 \nabla ^{\ast }f_{1} -\frac{\chi_S^{\prime}}{2}f_1\nabla^\ast f_2\;,
\label{eq:f2-SmoluBM08}
\end{align}
\end{subequations}
where $\nabla =\partial _{x}+i\partial _{y}$ and $\nabla ^{\ast }=\partial
_{x}-i\partial _{y}$ are the complex derivative operators.
Some of the notation has been changed as compared to the one used in  \cite{BaskaranPRL08}
to highlight the comparison with the equations obtained from the Boltzmann approach. Specifically, \eb{we have tried to keep notations as close as possible to that of \cite{Peshkov12}, using subscript 'S' or 'B' for the sets of continuum equations derived from the Smoluchowski and Boltzmann equations respectively. A few further coefficients needed to be added, like the coefficient $\lambda^{\prime }$ in Eq.~(\ref{eq:f1-SmoluBM08}), for which we kept the original notation since no similar term appear in \cite{Peshkov12} (see below). Additionally, we have used
$\zeta_S=\lambda$, $\kappa_S=3\lambda^{\prime\prime}/5$, $\kappa_S^{\prime}=\lambda^{\prime\prime}/48$, $\chi_S=\lambda^{\prime\prime}/24$, and
$\chi^{\prime}_S = 3\lambda^{\prime\prime}/5$
where each of the $\lambda$'s are proportional to the length of the rods and the square of the self-propulsion speed.} The parameter 
$\mu_S =4D_R (\rho/\rho_c-1)$ controls the stability of the uniform isotropic state, with $\rho_c$ a critical density that scales inversely with the square of the length of the rods and the square of the self-propulsion speed, $v_0$, i.e., longer rods and faster rods are destabilized at lower densities.

On the other hand, the equations found in \cite{Peshkov12} using the Boltzmann
approach are (note that in \cite{Peshkov12} $v_0$ was set to $1$)\footnote{As mentionned in \cite{Bertin13}, Eq.~(\ref{eq:f2-Peshkov}) as published in \cite{Peshkov12} includes a misprint, as the term $\frac{\nu}{4} \Delta f_2$ was erroneously written $\frac{\nu}{4} \nabla^2 f_2$, which is incorrect since complex operators are used here instead of vectorial ones.}
\begin{subequations}
\begin{align}
\partial_t \rho &+v_0 \mathrm{Re}(\nabla^* f_1 )=0\;,
\label{eq:rho-Peshkov}\\
\partial_t f_1 &= -(\alpha - \beta|f_2|^2)f_1+ \zeta_B f^*_1 f_2 -\frac{v_0}{2} (\nabla\rho + \nabla^* f_2)  \notag \\
& + \frac{\gamma}{2} f^*_2 \nabla f_2\;,
\label{eq:f1-Peshkov} \\
\partial_t f_2 &= (\mu_B-\xi |f_2|^2) f_2-\frac{v_0}{2} \nabla f_1 -\frac{\kappa_B}{2} f^*_1 \nabla f_2 \notag \\
& -\frac{\chi_B}{2} f_2 \nabla^* f_1  -\frac{\chi_B}{2} f_1 \nabla^* f_2 \notag \\
& +\frac{\nu}{4} \Delta f_2+\omega f_1^2 +\tau |f_1|^2f_2\;. 
\label{eq:f2-Peshkov}
\end{align}
\end{subequations}
where $\Delta=\nabla \nabla^{\ast}$ is the Laplacian.
Note that both coefficients $\mu_S$ and $\mu_B$ are positive at high
density and/or low noise, leading to a linear instability of the isotropic
state ($f_1=f_2=0$) towards the onset of nematic order ($f_2 \neq 0$).

In spite of a number of similarities, the sets of equations
(\ref{eq:rho-SmoluBM08},\ref{eq:f1-SmoluBM08},\ref{eq:f2-SmoluBM08})
and (\ref{eq:rho-Peshkov},\ref{eq:f1-Peshkov},\ref{eq:f2-Peshkov})
do exhibit some differences. We now highlight them and identify their origin, summarizing the
detailed analysis given in subsequent sections.

\begin{enumerate}
\item Equation~(\ref{eq:f2-SmoluBM08}) for $f_2$, obtained from the Smoluchowski equation, does not contain a saturating non-linear term  $|f_{2}|^{2}f_{2}$ that is needed to cutoff the linear instability and obtain an ordered state (although this term was added by hand in \cite{BaskaranPRL08}).  When positional diffusion is neglected, Eq.~(\ref{eq:f2-SmoluBM08}) does not contain a diffusion term $\Delta f_2$ either.
Both these terms are present in Eq.~(\ref{eq:f2-Peshkov}) thanks to the Ginzburg-Landau closure ansatz used in \cite{Peshkov12} that includes higher order modes than the closure used in \cite{BaskaranPRL08}.%
\footnote{
In \cite{BaskaranPRL08}, the fast modes $f_{3}$ and $f_{4}$ were discarded. This approach is in line with
standard practice in kinetic theory of molecular or granular gases, where a
closure relation is obtained by constraining the one-particle phase-space
distribution to be a function of the slowly relaxing fields (the so-called
normal form of the distribution) \cite{Garzo}.
A higher order closure is, however, needed in both passive and active systems to derive the nonlinear terms that yield the ordered phase (see \cite{Peshkov14} in the case of active systems). Note that this was recognized by Baskaran and Marchetti in \cite{BaskaranPRL08} where the nonlinear term $|f_{2}|^{2}f_{2}$ was added by hand.}
This closure also yields  terms $|f_2|^2 f_1$ and
$f_2^{\ast} \nabla f_2$ that are included in Eq.~(\ref{eq:f1-Peshkov}), but not in  Eq.~(\ref{eq:f1-SmoluBM08}).

\item The terms $f_{1}^{\ast }\nabla f_{1}$, $f_{1}\nabla ^{\ast }f_{1}$,
and $f_{1}\nabla f_{1}^{\ast }$ are present in Eq.~(\ref{eq:f1-SmoluBM08}),
but not in Eq.~(\ref{eq:f1-Peshkov}). 
Similarly, a term $f_{2}\nabla f_{1}^{\ast }$ is present in
Eq.~(\ref{eq:f2-SmoluBM08}), but not in Eq.~(\ref{eq:f2-Peshkov}).
These terms arise because the Smoluchowski equation derived in \cite{BaskaranPRL08} incorporates excluded volume forces and torques arising from the finite size of the particles. In contrast, the  Boltzmann equation used in Ref.~\cite{Peshkov12} considers effective alignment rules between point-like particles, and thus includes only torques (acting during collisions), but no  forces. We note that the finite size of particles could also be incorporated in a Boltzmann approach \cite{Patelli}.
This difference is therefore at the level of
the underlying microscopic model and is not associated with differences in the closures used.

\item The terms $f_{1}^{2}$ and $|f_{1}|^{2}f_{2}$ are present in
Eq.~(\ref{eq:f2-Peshkov}), but not in Eq.~(\ref{eq:f2-SmoluBM08}).
As we will show below, this difference is also due to the different microscopic models used by the two sets of authors, which results in a different symmetry of the collision kernels considered in the two kinetic equations. Baskaran and Marchetti  considered long, thin rods whose collisions are described by a kernel with pure nematic symmetry.  Peshkov {\em et al.} considered instead point particles with nematic alignment and an effective circular excluded volume, resulting in a collision kernel that contains both terms of nematic and polar Fourier components.
%In \cite{BaskaranPRL08}, nematic symmetry of the Smoluchowski torque kernel was obtained by assuming infinitely thin rods. In contrast, the kernel used in \cite{Peshkov12} describes point particles with prescribed aligning interactions and contains both nematic and polar Fourier components, due to the collision rate that has a polar symmetry.
We will see below that considering infinitely thin rods in the Boltzmann framework also leads to a kernel with nematic symmetry.

\end{enumerate}

In this section, we have summarized the differences between the two models
both at the kinetic and at the hydrodynamic level.
We have also briefly identified the origin of these differences.
The technical aspects of these conclusions are unfolded in detail in the
subsequent sections.
In section~\ref{sect-point-like}, we show that a strong formal analogy emerges between the Smoluchowski and Boltzmann equations when considering the limit of point-like particles with interactions reducing to alignment rules. The continuum equations obtained for the order parameters have essentially the same structure in both cases, and take precisely the same form in the limit of thin rods.
Then, in section \ref{sect-force}, we explain how further terms emerge when including forces in addition to alignment torques in the Smoluchowski equation, thus obtaining a generalization of Eqs.~(\ref{eq:f1-SmoluBM08},\ref{eq:f2-SmoluBM08}), taking into account the relevant (lowest order) nonlinear terms.

\section{Point-like particles with nematic alignment interactions}
\label{sect-point-like}

In this section, we derive continuum equations both from the Smoluchowski and Boltzmann equations
for point-like particles interacting only via alignment interactions (i.e., torques). In the case of the Smoluchowski equation, this corresponds to only incorporating the mean torque given by the first term in Eq.~\eqref{eq:tau}, but neglecting the mean forces given by the second term of Eq.~\eqref{eq:tau} and Eq.~\eqref{eq:F}.  In the case of the Boltzmann equation only torques were included from the outset in Eq.~\eqref{1.2}.
By point-like particles, we mean that we consider the limit where explicit excluded volume contributions are neglected, although
the particles still have a ``shape'' that defines the region of interaction and therefore determines their collision rate.  This shape is chosen as needle-like in the work by Baskaran and Marchetti and as circular in the work by Peshkov {\em et al.}.

\subsection{Smoluchowski equation}

\label{sec-smolu}

Following an approach similar to the one used \cite{Farrell,Grossmann13,Grossmann14} for polar self-propelled particles,
we first consider the force-free Smoluchowski equation, in which interaction
between particles appears through the average torque $\tau$ exerted
by neighboring particles (from now on, we set $v_0=1$)
\begin{equation}
\partial_t f + \mathbf{e}(\theta) \cdot {\bm \nabla} f = D_R
\partial_\theta^2 f - \partial_\theta (f \tau)\;.
\end{equation}
The local average torque $\tau$ can generically be expressed as
\begin{equation}  \label{def-tau}
\tau(\mathbf{r},\theta,t) = \int d\mathbf{r}^{\prime }\int d\theta^{\prime }%
\tilde{K}(\mathbf{r}^{\prime }-\mathbf{r},\theta^{\prime },\theta) f(\mathbf{%
r}^{\prime },\theta^{\prime },t)\;.
\end{equation}
The gradient term given in Eq.~\eqref{eq:tau} is obtained by expanding Eq.~\eqref{def-tau} in gradients and assuming that $|\mathbf{r}-\mathbf{r}'|$ is of order of the length of the rods. To simplify the comparison with the Boltzmann approach that considers point particles, we neglect these terms here and simply write
$\tilde{K}(\mathbf{r}^{\prime }-\mathbf{r},\theta^{\prime },\theta)=\delta(%
\mathbf{r}^{\prime }-\mathbf{r}) K(\theta^{\prime }-\theta)$, where we have
also taken into account rotational invariance. 
We assume space reversal symmetry, which leads to $K(-\theta)=-K(\theta)$.
We also assume that $K(\theta)$ obeys a nematic symmetry
$K(\theta+\pi)=K(\theta)$, which is valid for thin rods undergoing collisions, 
when cap-on-cap collisions can be neglected \cite{BaskaranPRL08}.
Hence the local average torque reads
\begin{equation}
\tau(\mathbf{r},\theta,t) = \int d\theta^{\prime }K(\theta^{\prime }-\theta)
f(\mathbf{r},\theta^{\prime },t)
\end{equation}

Passing to Fourier components, we obtain the following hierarchy of equations:
\begin{eqnarray}  \label{eq:fk-smolu0}
&& \partial_t f_k+\frac{1}{2}\nabla f_{k-1}+\frac{1}{2}\nabla^* f_{k+1} \\
&& \qquad \qquad \qquad = - D_k^{S} f_k + \frac{ik}{2\pi}\sum_{q=-\infty}^{%
\infty} \hat{K}_{-q} f_{k-q} f_q  \notag
\end{eqnarray}
where $D_k^{S}=D_R k^2$, and
\begin{equation}  \label{eq:Kq}
\hat{K}_q = \int_{-\pi}^{\pi} e^{iq\theta} K(\theta) d\theta
\end{equation}
Due to the nematic symmetry of the interaction, all odd Fourier modes $\hat{K%
}_{2m+1}=0$.
Defining
\begin{equation}
J_{k,q}^{S} = \frac{ik}{2\pi} \hat{K}_{-q}
\end{equation}
we get for the Fourier transform of the Smoluchowski equation
\begin{eqnarray}  \label{eq:fk-smolu1}
&& \partial_t f_k+\frac{1}{2}\nabla f_{k-1}+\frac{1}{2}\nabla^* f_{k+1} \\
&& \qquad \qquad \qquad = - D_k^{S} f_k + \sum_{q=-\infty}^{\infty}
J_{k,q}^{S} f_{k-q} f_q  \notag
\end{eqnarray}
Note that due to the symmetry $K(-\theta)=-K(\theta)$, $J_{k,q}^{S}$ is
real. In addition, $J_{k,q}^{S}=0$ for odd $q$ due to the nematic symmetry
of the interaction.

For $k=0$, Eq.~(\ref{eq:fk-smolu1}) directly leads to the continuity
equation
\begin{equation}
\partial_t \rho + \mathrm{Re}(\nabla^* f_1) = 0
\end{equation}

To derive closed equations for $f_1$ and $f_2$, we need to resort to an
approximation scheme. For an almost isotropic distribution, that is, for
small values of the Fourier harmonics $f_k$ ($k>1$), Eq.~(\ref{eq:fk-smolu1}%
) can be rewritten to linear order as
\begin{equation}  \label{eq:fk-smolu-lin}
\partial_t f_k+\frac{1}{2}\nabla f_{k-1}+\frac{1}{2}\nabla^* f_{k+1} =
\mu_k^{S} f_k
\end{equation}
with $\mu_k^{S} = (J_{k,k}^{S}+J_{k,0}^{S}) \rho - D_k^{S}$. At low density,
the linear coefficient $\mu_k^{S}$ is negative. Whether $\mu_k^{S}$ becomes
positive or not at higher density depends on the sign of $%
J_{k,k}^{S}+J_{k,0}^{S}$. The precise values of these coefficients
depend on the details of the chosen interactions, which we do not specify
explicitly here. We however assume that interactions favor nematic alignment,
resulting in $J_{2,2}^{S}+J_{2,0}^{S}>0$. As a result, $\mu_2^{S}$ becomes
positive above a transition density $\rho_t$ --or equivalently, below a
given threshold value of $D_2^{S}$ at fixed density. For $\rho$ just
slightly above $\rho_t$, $\mu_2^{S}$ is positive and small, leading to a
slow instability of the state $f_2=0$, a regime in which the dynamics of the
system can be reduced to that of a few coupled modes.
\eb{In constrast, we assume that $J_{1,1}^{S}+J_{1,0}^{S}<0$ since interactions do not favor polar order, so that $\mu_1^S$ remains negative.}

Following \cite{Peshkov12,Peshkov14}, we introduce a truncation procedure
close to the instability threshold of the linearized equation. We use the
following scaling ansatz, with $\epsilon$ a small parameter:
\begin{equation}
\rho-\rho_0 \sim f_1 \sim f_2 \sim \epsilon, \quad f_{2k-1} \sim f_{2k} \sim
\epsilon^k \; \; (k>0)
\end{equation}
To lighten notations, we further introduce the coefficient $C_{k,q}^{S}$
defined as
\begin{equation}
C_{k,q}^{S} = J_{k,q}^{S}+J_{k,k-q}^{S}
\end{equation}
The nematic symmetry, which implies $J_{k,q}^{S}=0$ for odd $q$, in turn
leads to $C_{k,q}^{S}=0$ if $k$ is even and $q$ is odd: $C_{2m,2l+1}^{S}=0$.

After truncation of Eq.~(\ref{eq:fk-smolu1}) to order $\epsilon^3$ for $k=1$
and $k=2$, one obtains the equations governing the evolution of $f_1$ and $%
f_2$,
\begin{subequations}
\begin{align}
\label{eq:f1-0}
\partial_t f_1 + \frac{1}{2}(\nabla^* f_2 + \nabla \rho) &= \mu_1^{S} f_1 \\
&+ C_{1,2}^{S} f_1^* f_2 + C_{1,3}^{S} f_2^* f_3  \notag \\
\partial_t f_2 + \frac{1}{2}(\nabla^* f_3 + \nabla f_1) &= \mu_2^{S} f_2 +
C_{2,4}^{S} f_2^* f_4  \label{eq:f2-0}
\end{align}
\end{subequations}
These equations are not closed, as they also involve the higher order
harmonics $f_3$ and $f_4$. We thus make use of Eq.~(\ref{eq:fk-smolu1}) for $%
k=3$ and $k=4$, truncating them to order $\epsilon^2$ since $f_3$ and $f_4$
appear in Eqs.~(\ref{eq:f1-0}) and (\ref{eq:f2-0}) only in space derivatives
or multiplied by another small field. We then obtain
\begin{subequations}
\begin{align}  \label{eq:f3}
f_3 &= \frac{1}{2\mu_3^{S}} \nabla f_2 - \frac{C_{3,2}^{S}}{\mu_3^{S}} f_1
f_2 \\
f_4 &= -\frac{C_{4,2}^{S}}{2\mu_4^{S}} f_2^2  \label{eq:f4}
\end{align}
\end{subequations}
Injecting Eqs.~(\ref{eq:f3}) and (\ref{eq:f4}) in Eqs.~(\ref{eq:f1-0}) and (%
\ref{eq:f2-0}), we obtain closed equations for $f_1$ and $f_2$,
\begin{subequations}
\begin{align}
\partial_t f_1 &= \mu_1^{S} f_1 -\frac{C_{1,3}^{S}\, C_{3,2}^{S}}{\mu_3^{S}}
|f_2|^2 f_1 + C_{1,2}^{S} f_1^* f_2  \notag \\
& \quad - \frac{1}{2} \nabla \rho - \frac{1}{2} \nabla^* f_2 + \frac{%
C_{1,3}^{S}}{2\mu_3^{S}} f_2^* \nabla f_2  \label{eq:f1-smolu} \\
\partial_t f_2 &= \mu_2^{S} f_2 - \frac{C_{2,4}^{S} C_{4,2}^{S}}{2\mu_4^{S}}
|f_2|^2 f_2 -\frac{1}{2} \nabla f_1  \notag \\
& \quad - \frac{1}{4\mu_3^{S}} \Delta f_2 + \frac{C_{3,2}^{S}}{2\mu_3^{S}}
\nabla^* (f_1 f_2)  \label{eq:f2-smolu}
\end{align}
\end{subequations}

\subsection{Boltzmann equation}

\label{sec-boltzmann}

In the case of the Boltzmann equation, interactions are described as
instantaneous collisions, and the equation reads

\begin{equation}  \label{Boltzmann-eq}
\partial_t f + \mathbf{e}(\theta)\cdot {\bm \nabla} f =
I_\mathrm{dif}[f] + I_\mathrm{col}[f]
\end{equation}
where $I_\mathrm{dif}$ and $I_\mathrm{col}$ are defined in Eqs.~(\ref{1.3}) and (\ref{1.4}) respectively.
The dynamics is defined such that the angles $\theta_1^{\prime }$ and $\theta_2^{\prime }$ after collisions are given by
\begin{equation}
\theta_1^{\prime }= \Psi(\theta_1,\theta_2) + \eta_1, \quad \theta_2^{\prime
}= \Psi(\theta_2,\theta_1) + \eta_2
\end{equation}
where $\eta_1$ and $\eta_2$ are independent noises drawn from $P(\eta)$. We
assume here that $\Psi(\theta_1,\theta_2)$ favors nematic alignment. Using
symmetry properties, $\Psi(\theta_1,\theta_2)$ can generically be
parameterized as \cite{Peshkov14}
\begin{equation}
\Psi(\theta_1,\theta_2) = \theta_1 + H(\theta_2 - \theta_1)
\end{equation}
\eb{where $H$ is an arbitrary function which encodes the nematic symmetry of the interaction, and is thus $\pi$-periodic.}
After expansion in angular Fourier series, one finds for the Boltzmann
equation \cite{Peshkov12,Peshkov14},
\begin{eqnarray}  \label{eq:fk-boltz}
&& \partial_t f_k+\frac{1}{2}\nabla f_{k-1}+\frac{1}{2}\nabla^* f_{k+1} \\
&& \qquad \qquad \qquad = - D_k^{B} f_k + \sum_{q=-\infty}^{\infty}
J_{k,q}^{B} f_{k-q} f_q  \notag
\end{eqnarray}
where the coefficients $D_k^{B}$ and $J_{k,q}^{B}$ are given by
\begin{subequations}
\begin{align}
D_k^{B} &= -\lambda_R (1-\hat{P}_k) \\
J_{k,q}^{B} &= \hat{P}_k I_{k,q} - I_{k,0}
\end{align}
\end{subequations}
with
\begin{subequations}
\begin{align}
\hat{P}_k &= \int_{-\infty}^{\infty} d\eta\, P(\eta) \, e^{ik\eta} \\
I_{k,q} &= \int_{-\pi}^{\pi} d\theta\, K_B(\theta)\, e^{-iq\theta +
ikH(\theta)}
\end{align}
\end{subequations}
Note that Eq.~(\ref{eq:fk-boltz}) is formally identical to Eq.~(\ref%
{eq:fk-smolu1}). The only difference, apart from the specific values of the
coefficients, is that $J_{k,q}^{B}$ is a priori nonzero for all $(k,q)$, as
it is not constrained by a nematic symmetry. In the explicit model
considered in \cite{Peshkov12} where particles are considered to have an
almost circular shape of diameter $d_0$ (the interaction radius), one has (setting $v_0=1$)
\begin{equation}  \label{eq:collision-rate}
K_B(\theta^{\prime }-\theta) = 4d_0 \left| \sin\frac{\theta^{\prime }-\theta%
}{2} \right|
\end{equation}
which is not invariant under the nematic symmetry $\theta \to \theta + \pi$.
This absence of nematic symmetry is due to the fact that $K(\theta^{\prime
}-\theta)$ corresponds to the collision rate of polar particles, which
depends on the velocity difference of the particles. However, for
non-circular particles, a simple heuristic generalization of
Eq.~(\ref{eq:collision-rate}) can be proposed, modulating the collision rate
by an orientation-dependent interaction radius $d(\theta^{\prime }-\theta)$.
In the limit of infinitely thin rods, one finds (in the frame moving with the
particle of orientation $\theta$),
\begin{equation}
d(\theta^{\prime }-\theta) = d_0 \left| \cos\frac{\theta^{\prime }-\theta}{2}
\right|
\end{equation}
so that the collision kernel obtained in this limit,
\begin{equation} \label{eq:Kbthin}
K_B^{\mathrm{thin}}(\theta^{\prime }-\theta) = 4d(\theta^{\prime }-\theta)
\left| \sin\frac{\theta^{\prime }-\theta}{2} \right| = 2d_0
|\sin(\theta^{\prime }-\theta)|
\end{equation}
indeed obeys a nematic symmetry. One thus recovers, in the limit of
infinitely thin rods, the property $J_{2m,2l+1}^{B}=0$, as in the
Smoluchowski case.

The derivation of the continuum equations for the polar and nematic order
parameters follows exactly the same lines as in Sect.~\ref{sec-smolu}, and
one finds (see \cite{Peshkov14} for details)
\begin{subequations}
\begin{align}
\partial_t f_1 &= \mu_1^{B} f_1 -\frac{C_{1,3}^{B}\, C_{3,2}^{B}}{\mu_3^{B}}
|f_2|^2 f_1 + C_{1,2}^{B} f_1^* f_2 \\
&- \frac{1}{2} \nabla \rho - \frac{1}{2} \nabla^* f_2 + \frac{C_{1,3}^{B}}{%
2\mu_3^{B}} f_2^* \nabla f_2  \notag \\
\partial_t f_2 &= \mu_2^{B} f_2 - \frac{C_{2,4}^{B} C_{4,2}^{B}}{2\mu_4^{B}}
|f_2|^2 f_2 \\
&+ \frac{1}{2} C_{2,1}^{B} f_1^2 - \frac{C_{2,3}^{B} C_{3,2}^{B}}{\mu_3^{B}}
|f_1|^2 f_2 -\frac{1}{2} \nabla f_1  \notag \\
&- \frac{1}{4\mu_3^{B}} \Delta f_2 + \frac{C_{3,2}^{B}}{2\mu_3^{B}}
\nabla^* (f_1 f_2) + \frac{C_{2,3}^{B}}{2\mu_3^{B}} f_1^* \nabla f_2  \notag
\end{align}
\end{subequations}
These equations have essentially the same form as Eqs.~(\ref{eq:f1-smolu})
and (\ref{eq:f2-smolu}). The additional terms $f_1^2$, $|f_1|^2 f_2$ and $%
f_1^* \nabla f_2$ that appear in the equation for $f_2$ are due to the lack of
nematic symmetry of the kernel $K_B(\theta^{\prime }-\theta)$. 
These terms vanish in the limit of infinitely thin rods, where one recovers the same equations as obtained by Baskaran and Marchetti on the basis of the Smoluchowski equation.
In other words, when only torques, but no forces are included in the kinetic equation, the two kinetic equations yield the same continuum equations and the differences in the published continuum equations obtained by the two approaches are entirely due to differences in the microscopic models.

\section{Particles with interaction forces and torques}
\label{sect-force}

%Up to now, we have mostly found that Smoluchowski and Boltzmann equations lead to similar continuum equations for the order parameters, with some terms which vanish in the Smoluchowski case due to the nematic symmetry. 
%However, this first result does not allow
%us to understand the more complicated equations (\ref{eq:rho-SmoluBM08}), (%
%\ref{eq:f1-SmoluBM08}) and (\ref{eq:f2-SmoluBM08}) obtained in \cite%
%{BaskaranPRL08}. 
To understand the remaining  differences between the two set of published continuum equations we need to examine the remaining terms in the Smoluchowski equation. Specifically, Eq.~(\ref{eq:f1-SmoluBM08}) contains terms
coupling the polar order parameter with its space derivative that are not in Eq.~\eqref{eq:f1-Peshkov}.
In terms of the complex notations used here, such terms
yield a linear combination of $f_1^* \nabla f_1$, $f_1 \nabla^* f_1$ and
$f_1 \nabla f_1^*$. 
%The appearance of these terms is surprising, given that
%similar terms were previously known to appear, in systems with polar
%interactions, due to a closure relation that expresses $f_2$ as a function
%of $f_1$, while we keep here both $f_1$ and $f_2$ as independent fields.
Also, the term $f_2 \nabla f_1^*$ appearing in Eq.~(\ref{eq:f2-SmoluBM08}) was not obtained in Eq.~(\ref{eq:f2-smolu}).

These additional terms arise because the Smoluchowski equation derived by Baskaran and Marchetti contains two additional terms as compared to the Boltzmann equation used by Peshkov {\em et al.}. The first one is the mean force given in Eq.~\eqref{eq:F} that describes momentum transfer in a collision. The second one is the second term in the torque given in Eq.~\eqref{eq:tau}, that arises from the difference in the position of the center of two colliding rods, and hence incorporates the finite size of the particles. Note that both terms could also be incorporated in a Boltzmann approach. Both terms are of first order in the spatial gradients and yield terms of the same symmetry in the continuum equations. In this section we discuss the terms arising form the mean force, while those due to the nonlocal torque are discussed in Appendix~\ref{appendix-A}.
%The main reason for the emergence of these additional terms in the
%continuum equations derived in \cite{BaskaranPRL08} is the presence of an
%average force term in the Smoluchowski equation, in addition to the average
%torque term considered in Sec.~\ref{sec-smolu}. Some of these terms can also
%be generated by taking into account the finite size of the particles, or in
%other words, by taking into account non-purely local interactions. In the
%present section, we however focus on purely local interactions, but we shall
%come back to the non-locality issue in the discussion section (see also Appendix~\ref{appendix-A}).
In addition, Baskaran and Marchetti incorporated positional diffusion in
\cite{BaskaranPRL08}, leading to the following Smoluchowski equation, with the short notation $\mathbf{e}=\mathbf{e}(\theta)$,
\begin{eqnarray}
\partial_t f + e_{\alpha} \partial_{\alpha} f &=& D_0 \Delta f + D_1
\left(e_{\alpha} e_{\beta}-\frac{1}{2}\delta_{\alpha\beta}\right)
\partial^2_{\alpha\beta} f  \notag \\
&+& D_R \partial_\theta^2 f - \partial_\theta (f \tau) - {\bm \nabla}\cdot
(f {\bm F})  \label{eq:full-smolu}
\end{eqnarray}
where $D_0$ and $D_1$ are the isotropic and anisotropic diffusion
coefficients, and $\mathbf{F}$ and $\tau$ are the force and torque given in Eqs.~\eqref{eq:F} and \eqref{eq:tau}, respectively (but we will ignore here the second term on the right hand side of Eq.~\eqref{eq:tau} and discuss it in Appendix~\ref{appendix-A}).

Starting from the Smoluchowski equation \eqref{eq:full-smolu}, the following continuum equations (rewritten here in the current complex notation) have been obtained \cite{BaskaranPRL08},
\begin{subequations}
\begin{align}
\partial_t \rho &+ {\rm Re} (\nabla^* f_1) = D_{\rho} \Delta \rho + \frac{D_Q}{2} {\rm Re} (\nabla^{*2} f_2)
\label{eq:rho-SmoluBM08-diff}\\
\partial_t f_1 &= -D_R f_1 -\frac{1}{2} \nabla \rho -\frac{1}{2} \nabla^* f_2 +\lambda f_1^* f_2 \notag\\
&- \lambda' \big( f_1^*\nabla f_1 + f_1\nabla^* f_1- f_1 \nabla f_1^* \big) \notag\\
&+ \frac{1}{2}(D_{b}+D_{spl}) \Delta f_1 + \frac{1}{2}(D_{b}-D_{spl}) \nabla^2 f_1^*
\label{eq:f1-SmoluBM08-diff}\\
\partial_t f_2 &= \mu_S f_2 -\frac{\tilde{v_0}}{2} \nabla f_1 + \frac{D_Q}{16} \nabla^2 \rho  \notag\\
& - \frac{3\lambda''}{10} (f_1^* \nabla f_2 + f_1 \nabla^* f_2) 
- \frac{\lambda''}{96} (2f_2 \nabla^* f_1 + f_2 \nabla f_1^*)
\label{eq:f2-SmoluBM08-diff}
\end{align}
\end{subequations}
where $\tilde{v}_0$ is a renormalized speed.
In the following, we aim at rederiving the generic form of Eqs.~(\ref{eq:rho-SmoluBM08-diff},\ref{eq:f1-SmoluBM08-diff},\ref{eq:f2-SmoluBM08-diff}), including the relevant nonlinear terms.
%Our starting point is the Smoluchowski equation, taking into account positional diffusion and the average force and torque exerted by neighboring particles.
%This equation reads, with the short notation $\mathbf{n}=\mathbf{e}(\theta)$,
%\begin{eqnarray}
%\partial_t f + n_{\alpha} \partial_{\alpha} f &=& D_0 \Delta f + D_1
%\left(n_{\alpha} n_{\beta}-\frac{1}{2}\delta_{\alpha\beta}\right)
%\partial^2_{\alpha\beta} f  \notag \\
%&+& D_R \partial_\theta^2 f - \partial_\theta (f \tau) - {\bm \nabla}\cdot
%(f {\bm F})  \label{eq:full-smolu}
%\end{eqnarray}
%where $D_0$ and $D_1$ are the isotropic and anisotropic diffusion
%coefficients, and where $\mathbf{F}$ is the average force exerted by
%neighboring particles, which can be expressed as
%\begin{equation}
%{\bm F}(\mathrm{r},\theta,t) = \int_{-\pi}^{\pi} d\theta^{\prime }\mathbf{G}%
%(\theta,\theta^{\prime }) f(\mathrm{r},\theta^{\prime },t).
%\end{equation}
%

On general grounds, the \eb{kernel ${\bm G}(\theta,\theta^{\prime })$ in Eq.~\eqref{eq:F} associated to the force} exerted by
a particle of orientation $\theta^{\prime }$ on a particle of orientation $%
\theta$ can be decomposed onto the directions parallel and perpendicular
to $\mathbf{e}(\theta)$,
\begin{equation}
\mathbf{G}(\theta,\theta^{\prime }) = G_{||}(\theta^{\prime }-\theta)\,
\mathbf{e}(\theta) + G_{\perp}(\theta^{\prime }-\theta)\, \mathbf{e}%
_{\perp}(\theta)
\end{equation}
with $\mathbf{e}_{\perp}(\theta)=\mathbf{e}(\theta+\frac{\pi}{2})$. The
scalar functions $G_{||}$ and $G_{\perp}$ depend only on the angle
difference, by rotational symmetry.
For non-chiral particles, the force ${\bm G}(\theta,\theta^{\prime })$ obeys a reflection symmetry, characterized by
\begin{equation}  \label{eq:symG-refl}
G_{||}(-\theta) = G_{||}(\theta), \quad G_{\perp}(-\theta) =
-G_{\perp}(\theta)
\end{equation}
The angular Fourier transform of $f{\bm F}$ can be decomposed into parallel and transverse contributions:
\begin{equation}
\int_{-\pi}^{\pi} d\theta\, e^{ik\theta} f(\mathbf{r},\theta,t) {\bm F}(\mathbf{r},\theta,t)
 = {\bm F}_k^{||} + {\bm F}_k^{\perp}
\end{equation}
with
\begin{subequations}
\begin{align}
\!\!\!\!\! 
{\bm F}_k^{||} &= \int_{-\pi}^{\pi} d\theta e^{ik\theta}
f(\theta) \int_{-\pi}^{\pi} d\theta^{\prime }f(\theta^{\prime })
G_{||}(\theta^{\prime }-\theta)\, \mathbf{e}(\theta) \\
\!\!\!\!\! 
{\bm F}_k^{\perp} &= \int_{-\pi}^{\pi} d\theta
e^{ik\theta} f(\theta) \int_{-\pi}^{\pi} d\theta^{\prime }f(\theta^{\prime
}) G_{\perp}(\theta^{\prime }-\theta) \mathbf{e}_{\perp}(\theta)
\end{align}
\end{subequations}
where we have written simply $f(\theta)$ instead of $f(\mathbf{r},\theta,t)$
to lighten notations. After some algebra, we can eventually write ${\bm %
\nabla}\cdot (f {\bm F})$ in the form
\begin{eqnarray}
{\bm \nabla}\cdot (f {\bm F}) &=& \sum_{q=-\infty}^{\infty} M_{-q} \nabla^* (f_q
f_{k+1-q})  \notag \\
&& \qquad\qquad + \sum_{q=-\infty}^{\infty} M_q \nabla (f_q f_{k-1-q})
\label{eq:div-F}
\end{eqnarray}
where
\begin{equation}
M_q = \frac{1}{4\pi} (\hat{G}_{q}^{||} + i \hat{G}_{q}^{\perp})
\end{equation}
with $\hat{G}_{q}^{||}$ and $\hat{G}_{q}^{\perp}$ the Fourier transforms of
$G_{||}$ and $G_{\perp}$
\begin{subequations}
\begin{align}
\hat{G}_{q}^{||} &= \int_{-\pi}^{\pi} d\theta \, e^{ik\theta} G_{||}(\theta) \\
\hat{G}_{q}^{\perp} &= \int_{-\pi}^{\pi} d\theta \, e^{ik\theta}
G_{\perp}(\theta)
\end{align}
\end{subequations}
The Fourier coefficients $\hat{G}_{q}^{||}$ and $\hat{G}_{q}^{\perp}$ are
constrained by symmetries.
The coefficient $\hat{G}_{q}^{||}$ is real and $\hat{G}_{q}^{\perp}$ is
purely imaginary due to the reflection symmetry Eq.~(\ref{eq:symG-refl}). It
thus follows that $M_q$ is real.
Note also that $\hat{G}_{-q}^{||}=\hat{G}_{q}^{||}$ and $\hat{G}%
_{-q}^{\perp}=-\hat{G}_{q}^{\perp}$, a property that we have used in Eq.~(%
\ref{eq:div-F}).

In Fourier transform, the Smoluchowski equation thus reads (see
\cite{Peshkov14} for a similar derivation of the anisotropic diffusion terms)
\begin{widetext}
\begin{align}
\partial_t f_k + \frac{1}{2} \nabla f_{k-1} + \frac{1}{2} \nabla^* f_{k+1}
&= D_0 \Delta f_k + \frac{D_1}{4} (\nabla^2 f_{k-2} + \nabla^{*2}
f_{k+2}) - D_k^{S} f_k + \sum_{q=-\infty}^{\infty} J_{k,q}^{S} f_{k-q} f_q \notag \\
&+ \sum_{q} M_{-q} \nabla^* (f_q f_{k+1-q})
+ \sum_{q} M_q \nabla (f_q f_{k-1-q})
\label{eq:full-smolu-Fourier}
\end{align}
\end{widetext}
As we show below, applying the closure scheme used in Ref.~\cite{BaskaranPRL08} one obtains continuum equations
for $\rho$, $f_1$ and $f_2$ that have the same form as those of
\cite{BaskaranPRL08}. Of course, the detailed expression of the coefficients
differs since we have neglected the gradient term in the torque, but we were interested here only in the structure of the continuum
equations.

To examine the effect of the closure, we truncate Eq.~(\ref{eq:full-smolu-Fourier}) to order $\epsilon^3$ for $\rho$%
, $f_1$ and $f_2$, yielding
\begin{widetext}
\begin{align}
&\partial_t \rho + \mathrm{Re}(\nabla^* f_1) = D_0 \Delta \rho +
\frac{D_1}{2} \mathrm{Re}(\nabla^{*2} f_2) + 2(M_0+M_{-1}) \mathrm{Re} \big(\nabla^* (\rho f_1) \big)
+ 2(M_{1}+M_{-2}) \mathrm{Re} \big(\nabla^* (f_1^* f_2) \big)
\label{eq:rho-fullSmolu} \\
&\partial_t f_1 + \frac{1}{2} \nabla^* f_2 + \big( \frac{1}{2}-M_0 \rho %
\big)\nabla \rho = \mu_1^{S} f_1 + D_0 \Delta f_1 + \frac{D_1}{4}\nabla^2 f_1^*
+ C_{1,2}^{S} f_1^* f_2 + C_{1,3}^{S} f_2^* f_3  \notag \\
& \qquad\qquad\qquad\qquad\qquad\qquad
+ (M_0+M_{-2}) \nabla^* (\rho f_2) + (M_2+M_{-2}) \nabla
|f_2|^2  + M_{-1} \nabla^* f_1^2 + (M_1+M_{-1}) \nabla |f_1|^2
\label{eq:f1-0-fullSmolu} \\
&\partial_t f_2 + \frac{1}{2} (\nabla^* f_3 + \nabla f_1) = \mu_2^{S} f_2 + D_0 \Delta f_2 + \frac{D_1}{4}\nabla^2 \rho + C_{2,4}^{S} f_2^* f_4 + (M_{-1}+M_{-2}) \nabla^* (f_1 f_2) \notag \\
& \qquad\qquad\qquad\qquad\qquad\qquad
+ (M_{-1}+M_{2}) \nabla(f_1^* f_2)  + (M_0+M_{-3}) \nabla^* (\rho f_3) + (M_0+M_{1}) \nabla (\rho
f_1)  \label{eq:f2-0-fullSmolu}
\end{align}
\end{widetext}
For $f_3$, Eq.~(\ref{eq:f3}) is changed into
\begin{equation}
f_3 = \frac{1}{\mu_3^{S}} \left(\frac{1}{2} -(M_0+M_2)\rho \right) \nabla
f_2 - \frac{C_{3,2}^{S}}{\mu_3^{S}} f_1 f_2
\end{equation}
while one recovers Eq.~(\ref{eq:f4}) for $f_4$. Using these last equations as
closure relations, we find for $f_1$ and for $f_2$
\begin{widetext}
\begin{subequations}
\begin{align}
& \partial_t f_1 = \mu_1^{S} f_1 - \Big(\frac{1}{2}-M_0\rho\Big) \nabla
\rho - \frac{1}{2} \nabla^* f_2 + D_0 \Delta f_1 
+ \frac{D_1}{4}\nabla^2 f_1^* -\frac{C_{1,3}^{S}\, C_{3,2}^{S}}{\mu_3^{S}}
|f_2|^2 f_1 + C_{1,2}^{S} f_1^* f_2 + \gamma f_2^* \nabla f_2 \notag\\
& \qquad \qquad \qquad
+ 2M_{-1} f_1\nabla^* f_1 + (M_1+M_{-1}) (f_1 \nabla f_1^* + f_1^*
\nabla f_1)
+ (M_0+M_{-2}) \nabla^* (\rho f_2) + (M_2+M_{-2}) f_2 \nabla f_2^*
\label{eq:f1-fullSmolu}\\
&\partial_t f_2 = \mu_2^{S} f_2 - \frac{C_{2,4}^{S} C_{4,2}^{S}}{2\mu_4^{S}}
|f_2|^2 f_2 -\frac{1}{2} \nabla f_1 
+ (M_0+M_1) \nabla (\rho f_1) + D \Delta f_2 + \frac{D_1}{4}
\nabla^2 \rho  \notag \\
& \qquad \qquad \qquad \qquad \qquad \qquad \qquad 
+ \chi (f_1\nabla^* f_2 + f_2 \nabla^* f_1)
+ (M_{-1}+M_2) (f_1^* \nabla f_2 + f_2\nabla f_1^*)
\label{eq:f2-fullSmolu}
\end{align}
\end{subequations}
\end{widetext}
where the coefficients are given by
\begin{subequations}
\begin{align}
\gamma &= \frac{C_{1,3}^{S}}{\mu_3^{S}} \Big(\frac{1}{2}%
-(M_0+M_2)\rho\Big)+(M_2+M_{-2})  \\
D &= D_0 - \frac{1}{\mu_3} \Big(\frac{1}{2}-(M_0+M_{-3})\rho\Big)
\Big(\frac{1}{2} -(M_0+M_2)\rho \Big)  \\
\chi &= \frac{C_{3,2}^{S}}{\mu_3^{S}}\Big( \frac{1}{2} - (M_0+M_{-3})\rho
\Big) + (M_{-1}+M_{-2}) 
\end{align}
\end{subequations}
Note that, unlike the torque, the force kernel used by Baskaran and Marchetti \cite{BaskaranPRL08} does not have the nematic symmetry, as in general ${\bm G}(\theta+\pi,\theta^{\prime })\not={\bm G}(\theta,\theta^{\prime })$. Were this \eb{symmetry present} the Fourier coefficients would satisfy $\hat{G}_{2n}^{||}=\hat{G}_{2n}^{\perp}=0$, leading to $M_{2n}=0$ for any integer $n$.

\section{Discussion}

In this paper, we have examined the differences between the continuum equations for interacting self-propelled rods previously obtained in the literature from the Smoluchowski equation  \cite{BaskaranPRL08} and from the Boltzmann equation in \cite{Peshkov12}. 
%We have shown that the differences arise either from (i) the use of different microscopic models or (ii) the use of different closures of the moment expansion of the kinetic equation. 
\eb{The differences, not surprisingly, arise either from (i) the use of different microscopic models or (ii) the use of different closures of the moment expansion of the kinetic equation.}
Concerning the model: Baskaran and Marchetti considered long, thin rods with finite and anisotropic excluded volume (hence incorporating momentum transfer in a collision and nonlocality  on the scale of the difference in position of the colliding rods), while Peshkov {\em et al.} considered point-like particles with nematic alignment rules and circular interaction areas. Additionally, while Baskaran and Marchetti used a simple truncation that neglects all moments higher than the second, Peshkov {\em  et al.} employed a more sophisticated closure that allows one to derive in particular the nonlinear term responsible for the onset of the ordered state.  We show here that when the same microscopic model and the same closure are used, the Smoluchowski and Boltzmann approach yield the same continuum equations, albeit with different microscopic expression for the parameters. This is perhaps not surprising, but it is reassuring to demonstrate the equivalence for these nonequilibrium systems.
%different closures of the kinetic theory.  can be understood from differences in (i) the symmetry of particle shape, (ii) the finite extension or not of the particles, (iii) the absence in \cite{BaskaranPRL08} of a closure relation involving higher modes.

One interesting result is that even in the simplest case of point-like particles with nematic aligning interactions, where only strictly local aligning torques are included,  the shape of the interaction region controls the symmetry of the collision frequency and the symmetry of the interaction kernel. The needle-like particles considered by Baskaran and Marchetti yield a collision kernel that contains only even Fourier components in the angle describing the difference in orientation of the interacting particles, and hence has pure nematic symmetry. The circular point-particles considered by Peshkov {\em et al.} yield a collision kernel that contains both even and odd Fourier components, and hence has mixed polar and nematic symmetry. This leads to additional terms in the equations of Ref.~\cite{Peshkov12} not obtained in Ref.~\cite{BaskaranPRL08,BaskaranJSM10}. We expect that these additional terms will generally be present when one considers finite-thickness rods, so that cap-on-cap collisions are not negligible. Numerical integration however indicates that at low density, these terms do not play an important role in the behavior of the equation.

%We have seen that considering purely local alignment interactions, both the Smoluchowski and Boltzmann equations yield the same continuum equations for the polar and nematic order parameters, under the assumption of a nematic kernel resulting from infinitely thin rods --see \cite{BaskaranJSM10} and Eq.~(\ref{eq:Kbthin}) above.
%By introducing in addition the average force in the Smoluchowski equation, one basically
%generates (without taking explicitly into account the specific collision rules) all the terms present in Eqs.~(\ref{eq:rho-SmoluBM08}),
%(\ref{eq:f1-SmoluBM08}) and (\ref{eq:f2-SmoluBM08}).

The terms $f_1^* \nabla f_1$, $f_1 \nabla^*
f_1$ and $f_1 \nabla f_1^*$ present in the work by Baskaran and Marchetti, but not in that of Peshkov {\em et al.}, 
arise from both the mean force given in Eq.~\eqref{eq:F} and the nonlocal contribution to the torque in Eq.~\eqref{eq:tau}. 
Here the discussion of the  latter has been relegated to Appendix~A, but
both contributions to the kinetic equation must be included to consistently evaluate the coefficients of these terms, as shown in \cite{BaskaranPRL08}.
These terms encode the fact that the polarization in the hard rod models is actually the physical flow velocity of the fluid of rods. They include the convection term $\mathbf{P}\cdot\bm\nabla\mathbf{P}$ that is the analog of the convective nonlinearity in the Navier-Stokes equation of passive fluids and a $\bm\nabla|\mathbf{P}|^2$ contribution to the pressure that arises from self propulsion.
The mean force and nonlocal torque also yield
additional terms $\mathrm{Re}
\big(\nabla^* (\rho f_1) \big)$ and $\mathrm{Re} \big(\nabla^* (f_1^* f_2) %
\big)$ in the continuity equation (\ref{eq:rho-fullSmolu}). These
 do not appear in Eq.~(\ref{eq:rho-SmoluBM08}) because 
nonlinear terms in the hydrodynamic variables  in the continuity equation were neglected in  \cite{BaskaranPRL08}.
Isotropic and anisotropic diffusion terms, as appearing in
Eqs.~(\ref{eq:rho-SmoluBM08-diff}),
(\ref{eq:f1-SmoluBM08-diff}) and (\ref{eq:f2-SmoluBM08-diff}),
are also reproduced when positional diffusion is taken into account,
see Eqs.~(\ref{eq:rho-fullSmolu}), (\ref{eq:f1-fullSmolu})
and (\ref{eq:f2-fullSmolu}).
%however that the relative weight of their coefficient is different from that
%of Eq.~(\ref{eq:f1-SmoluBM08}). The same is true also for the coefficients
%of the terms $f_1\nabla^* f_2$, $f_2 \nabla^* f_1$, $f_1^* \nabla f_2$, $%
%f_2\nabla f_1^*$, when comparing Eq.~(\ref{eq:f2-fullSmolu}) with
%Eq.~(\ref{eq:f2-SmoluBM08}).
%One of the reasons explaining this discrepancy is that
%in \cite{BaskaranPRL08}, the finite size of particles has been taken into
%account explicitly, yielding slightly non-local interactions. Taking this
%correction into account yields new contributions of the same form as the
%existing terms (see Appendix~\ref{appendix-A}).
%It thus mainly amounts to a renormalization of the coefficients.
Finally, a renormalization of the velocity $v_0=1$,
in agreement with Eq.~(\ref{eq:f2-SmoluBM08-diff}), is obtained in Eq.~(\ref{eq:f2-fullSmolu}), by expanding the term $\nabla (\rho f_1)$.

Note that the non-locality of the interactions, 
%mostly affects the
%calculation of the average torque. We provide in Appendix~A a brief account
%of this extension. Since the average force $\mathbf{F}$ appears in a
%divergence in Eq.~(\ref{eq:full-smolu}), taking into account the
%non-locality of $\mathbf{F}$ mainly 
only contributes to order $%
\epsilon^4$ to the average force. These terms are therefore neglected in the truncation procedure. The only terms
of order $\epsilon^3$ in the equation for $f_k$ ($k=0$, $1$, $2$) are of the
form $\rho\Delta f_k$, $\rho \nabla^2 f_{k-2}$ and $\rho \nabla^{*2} f_{k+2}$ (for
$k=0$ only in this last case) and thus would just renormalize the existing
diffusion terms.

Finally, let us mention that the non-locality of interactions
(or, in other words, the finite excluded volume of particles)
can also be accounted for in the Boltzmann framework.
The corresponding study goes beyond the scope of the present note, and will
be the subject of a future publication \cite{Patelli}.

\section*{Acknowledgements}
MCM was supported by the National Science Foundation (NSF) award DMR-1305184 and by the Simons Foundation. AB acknowledges support from NSF-DMR-1149266, the Brandeis-MRSEC through NSF DMR-0820492 and NSF MRSEC-1206146. All authors thank the KITP at the University of California, Santa Barbara, for its hospitality during the initial stages of this work and support through NSF award PHY11-25915 and the Gordon and Betty Moore Foundation Grant No.~2919.

\appendix

\section{Non-local average torque}
\label{appendix-A}

In this appendix, we briefly sketch the derivation of the Fourier transform
of the Smoluchowski equation in the case of weakly non-local interactions
affecting the torque only. We start from the non-local average torque Eq.~(%
\ref{def-tau}), and first rewrite the kernel using the change of variables $%
\mathbf{r^{\prime }}-\mathbf{r}=s\mathbf{e}(\phi)$. From rotational
symmetry, we obtain
\begin{equation}
\tilde{K}(\mathbf{r^{\prime }}-\mathbf{r},\theta^{\prime },\theta) =
K(s,\theta^{\prime }-\theta,\phi-\theta)
\end{equation}
Expanding the torque to first order in gradient, we get
\begin{eqnarray}
\tau(\theta) &=& \int_{-\pi}^{\pi} d\theta^{\prime} \, K_1(\theta^{\prime }-\theta)
f(\theta^{\prime })  \notag \\
&+& \int_{-\pi}^{\pi} d\theta^{\prime} \int_{-\pi}^{\pi} d\phi\, K_2(\theta^{\prime
}-\theta, \phi-\theta)\, \mathbf{e}(\phi) \cdot \mathbf{\nabla} f(\theta^{\prime })
\notag \\
&\equiv& \tau_1(\theta) + \tau_2(\theta)
\end{eqnarray}
where we have defined
\begin{subequations}
\begin{align}
& \!\!\!\! K_1(\theta^{\prime }-\theta) = \int_{-\pi}^{\pi} d\phi
\int_0^{\infty} ds \,s K(s,\theta^{\prime }-\theta,\phi-\theta) \\
& \!\!\!\! K_2(\theta^{\prime }-\theta,\phi-\theta) = \int_0^{\infty} ds
\,s^2 K(s,\theta^{\prime }-\theta,\phi-\theta)
\end{align}
\end{subequations}
Note that Eq.~(\ref{eq:tau}) is recovered by introducing
\begin{equation}
\mathbf{K}_2(\theta^{\prime},\theta) = \int_{-\pi}^{\pi} d\phi\, K_2(\theta^{\prime
}-\theta, \phi-\theta)\, \mathbf{e}(\phi) \;.
\end{equation}
From Sect.~\ref{sec-smolu}, we know that the Fourier transform of $%
\tau_1(\theta)$ is given by
\begin{equation}
\int_{-\pi}^{\pi} d\theta\, e^{ik\theta} \tau_1(\theta) =
\sum_{q=-\infty}^{\infty} J_{k,q}^{S} f_{k-q} f_q
\end{equation}
A similar calculation for the Fourier transform of $\tau_2$ yields
\begin{equation}
\int_{-\pi}^{\pi} d\theta e^{ik\theta} \tau_2(\theta) = \sum_q \left( L_{-q} f_{k+1-q} \nabla^* f_q - L_{q} f_{k-1-q} \nabla f_q \right)
\label{eq:tau2}
\end{equation}
with %$L_q=i \hat{K}_{q,1}/(4\pi)$
\begin{equation}
L_q=\frac{i}{4\pi}\, \hat{K}_{q,1}
\end{equation}
and
\begin{equation}
\hat{K}_{q_1,q_2} = \int_{-\pi}^{\pi} d\theta_1 \int_{-\pi}^{\pi} d\theta_2
K_2(\theta_1,\theta_2)\, e^{-iq_1\theta_1-iq_2\theta_2}
\end{equation}
Note that due to the nematic symmetry $K_2(\theta_1+\pi,\theta_2+\pi)=K_2(%
\theta_1,\theta_2)$, one has $L_{2n}=0$, hence the sum may be carried over odd $q$'s only
in Eq.~(\ref{eq:tau2}). Pluging the expression (\ref{eq:tau2}) of the torque
into the Fourier transform of the Smoluchowski equation (with force term)
eventually leads to
\begin{eqnarray}
&& \partial_t f_k + \frac{1}{2} \nabla f_{k-1} + \frac{1}{2} \nabla^* f_{k+1}
\notag \\
&& \qquad = D_0 \Delta f_k + \frac{D_1}{4} (\nabla^2 f_{k-2} + \nabla^{*2}
f_{k+2})  \notag \\
&& \qquad - D_k^{S} f_k + \sum_{q=-\infty}^{\infty} J_{k,q}^{S} f_{k-q} f_q
\notag \\
&& \qquad + \sum_{q} (kL_{-q} + M_{-q} + M_{q-k-1}) f_{k+1-q} \nabla^* f_q
\notag \\
&& \qquad + \sum_{q} (kL_q + M_q +M_{k-1-q}) f_{k-1-q} \nabla f_q
\label{eq:NL-smolu-Fourier}
\end{eqnarray}
The resulting equation thus has the same structure as Eq.~(\ref%
{eq:full-smolu-Fourier}), after expansion of the derivatives of products and
relabelling. Only the values of the coefficients differ. This however leads
to a reweighting of the different terms which generically breaks the
symmetry between terms like $f_1\nabla^* f_2$ and $f_2 \nabla^* f_1$, or $%
f_1^* \nabla f_2$ and $f_2\nabla f_1^*$ observed in Eq.~(\ref%
{eq:f2-fullSmolu}).

\end{document}